\documentclass[prd,preprint,showpacs]{revtex4}
\usepackage{chay,epsfig}
\begin{document}
\title{Collinear effective theory at subleading order \\
and its application to heavy-light currents}
\author{Junegone Chay}
\email[e-mail address: ]{chay@korea.ac.kr}
\author{Chul Kim}
\affiliation{Department of Physics, Korea University, Seoul 136-701,
Korea} 
\preprint{KUPT--02--01}
\begin{abstract}
We consider a collinear effective theory of highly energetic quarks
with energy $E$, interacting with collinear and soft gluons by
integrating out collinear degrees of freedom to subleading order.
The collinear effective theory offers a systematic expansion in power 
series of a small parameter $\lambda \sim p_{\perp}/E$, where
$p_{\perp}$ is the transverse momentum of a collinear particle. We 
construct the effective Lagrangian to first order in $\lambda$, and
discuss its features including additional symmetries such as 
collinear gauge invariance and reparameterization invariance.
Heavy-light currents can be matched from the full
theory onto the operators in the collinear effective theory at one
loop and to order $\lambda$. We obtain heavy-light current operators
in the effective theory, calculate their Wilson
coefficients at this order, and the renormalization group equations
for the Wilson coefficients are solved. As an application,
we calculate the form factors for decays of $B$ mesons
to light energetic mesons to order $\lambda$ and at
leading-logarithmic order in $\alpha_s$.  
\end{abstract}

\pacs{13.25.Hw, 11.10.Hi, 12.38.Bx, 11.40.-q}

\maketitle

\section{\label{intro}Introduction}
When a $B$ meson decays into light mesons, we can explore different
kinematic regions depending on the momenta carried by the light
mesons. When a light meson is emitted from a heavy quark with momentum 
of order $\Lambda_{\mathrm{QCD}}$, this decay can be successfully
described by the heavy quark effective theory (HQET) \cite{hqet}. The
momentum of a heavy quark can be decomposed as $p_b = m_b v +k$, where
$k$ is the residual momentum of order $\Lambda_{\mathrm{QCD}}$. The
leading contribution to the decay corresponds to the
partonic result, and the corrections can be systematically
expanded in power series of $1/m_b$ and $\alpha_s$. Inclusive decays
of heavy mesons with large momentum transfer can be treated in the
HQET with the operator product expansion \cite{chay}. Exclusive decays
with heavy-heavy currents and heavy-light currents can also be treated
in the context of the HQET \cite{falk,neubert}. 

If a light meson from $B$ decays carries a large energy, HQET
alone is no longer useful since the large energy of a light meson can
be as large as $m_b$. Then an expansion
in $1/m_b$ alone is not appropriate. In this case, however, we can
construct a different type of an effective theory by taking the energy
$E$ of the energetic light quark to infinity. In this limit,
nonperturbative effects can also be systematically obtained. In
fact, this effective theory is more complicated than the HQET and the
naive power counting in $1/E$ should be modified since the system
involves several energy scales. 

Another complication arises in decays of a heavy quark with an
energetic light quark due a Sudakov logarithm since
there are both collinear and infrared divergences
\cite{sudakov}. There has been some discussion of summing Sudakov
logarithms using effective field 
theories \cite{grozin,aglietti, corbo, mannel}. Such an approach has
an advantage over conventional methods since effective theories are  
valid beyond perturbation theory, and it is straightforward to go
beyond the leading approximation by including higher-dimensional
operators. The main advantage of using
effective theories in this case is that we can reproduce the Sudakov
logarithm easily without dividing all the kinematic regions
\cite{smirnov}, and the 
calculation is manifest in the calculational procedure. However, we
need an effective theory in which logarithms arising at one loop in
the effective theory should match logarithms arising at one loop in
QCD for any matching scale $\mu$ in the minimal subtraction
scheme. Only in this case, these logarithms may be summed using the
renormalization group equations. The large-energy effective theory
suggested by Dugan and Grinstein \cite{dugan} does not satisfy this
criterion since it does not include the effects of collinear gluons
properly. 

Recently Bauer et al. \cite{luke} have proposed a new effective
theory called the ``collinear-soft effective theory''. If a light
quark moves with a large energy, the momentum has three distinct
scales. The momentum component in the light cone direction $n^{\mu}$
is the largest, of the order of the energy of the quark, $E$. The
transverse momentum is smaller than $E$, and the momentum component
opposite to the light cone direction is the smallest. 
In order to disentangle the three scales conveniently, a small
parameter $\lambda$ is introduced. The 
largest component has the momentum of order $E$. The transverse
component is of order $E\lambda$, and the smallest component is of
order $E\lambda^2$. 

Between $E$ and $E\lambda$, we have collinear
modes and soft modes for the light quark. Here we integrate out all
the collinear modes above some scale $\mu$, and the result is the
effective theory consisting of collinear quarks and soft
quarks. The effective theory
at this stage is called the collinear-soft effective theory,
which we will call the ``collinear  
effective theory'' for brevity. Below the scale $E\lambda$ and above
$E\lambda^2$, we integrate out all the collinear modes, and there
remain only soft modes in the final soft effective theory. This
actually corresponds to the large-energy 
effective theory suggested by Dugan and Grinstein \cite{dugan}, in
which there are only soft modes. In Ref.~\cite{luke}, they show
that at each stage of the effective theories, the infrared behavior of
the full theory is correctly reproduced by including the effects of
collinear gluons. Therefore heavy-light currents in the full theory
finally can be matched onto operators in the effective theories, their
Wilson coefficients are calculable and the renormalization group
equation can be solved. 

If we consider exclusive $B$ decays via heavy-light currents in the
scheme of effective theories, it is sufficient to consider the
collinear effective theory between the scale $E$ and $E\lambda$ and
integrate out all the degrees of freedom above some scale $\mu$. At
this scale, we describe a heavy quark in terms of HQET, and treat an
energetic light quark in the collinear effective theory. This limit
corresponds to $m_b, E \rightarrow \infty$ with $E/m_b$ fixed. We can
calculate the Wilson coefficients of various operators in the
effective theory by matching to the full theory and can obtain
anomalous dimensions of various operators. In this paper, we extend
further the idea of the collinear effective theory and derive the
effective Lagrangian to subleading order in $\lambda$ and renormalize
the effective theory at one loop. Also we consider the correction to 
heavy-light currents to order $\lambda$ and to leading logarithmic
order in $\alpha_s$.

In Section \ref{cet}, we briefly review the
collinear effective theory, and  derive the effective Lagrangian to
order $\lambda$. We also discuss a collinear gauge invariance in the
effective theory. In Section \ref{repinvar}, we discuss
reparameterization invariance in the collinear effective theory. The
reparameterization invariance ensures that the kinetic energy term
is not renormalized to all orders in $\alpha_s$. It is also useful in  
deriving high-dimensional operators for heavy-light currents in the
collinear effective theory and in obtaining the Wilson coefficients
and the renormalization behavior of these high-dimensional
operators. In Section \ref{match}, we match heavy-light currents
between the full QCD and the collinear effective theory, and consider
the effects of radiative corrections at one loop. In Section \ref{rg},
we compute the anomalous dimensions of various heavy-to-light
operators to order $\lambda$ at one loop, and solve the
renormalization group equation for the Wilson coefficients in the
collinear effective theory. In Section \ref{form}, we  
consider form factors of heavy-light currents for the vector and the
axial vector currents to order $\lambda$. In Section \ref{conc}, we
present a conclusion and perspectives of the collinear effective
theory. In Appendix, we present an explit calculation to show that the
effective Lagrangian at order $\lambda$ is not renormalized at one
loop.

\section{\label{cet}Collinear effective theory}
We construct an effective theory which describes the dynamics of
energetic light quarks. A detailed derivation of the effective
theory at leading order in $\lambda$ is described in Refs.~\cite{luke,
bauer1, bauer2, bauer3}, and we will briefly review the idea. Then we 
construct the effective theory to order $\lambda$. 
Let us consider a reference frame in which a light quark carries a
large energy $E$. If we neglect the quark mass, the only large
parameter in this system is the energy $E$ itself. 
Since we are interested in decays of heavy
mesons to energetic light hadrons, we can conveniently choose a
reference frame as the rest frame of a heavy meson, in which the
energy of light hadrons is indeed large in the heavy quark limit. In
this reference frame, light particles lie close on the light-cone
direction $n^{\mu}$, and we describe their dynamics using the
light-cone variables $p=(p^+, p^-, p_{\perp})$, where $p^+ = n\cdot
p$, and $p^- = \overline{n} \cdot p$. We choose the axis such that
$n^{\mu} = (1,0,0,1)$, $\overline{n}^{\mu} = (1,0,0,-1)$ with $n\cdot
\overline{n}=2$. 

For the energetic quark, there are three distinct energy scales, with
$p^- \sim 2E$ being large, while $p_{\perp}$ and $p^+$ are small. If
we take a small parameter as $\lambda \sim p_{\perp}/p^-$, we can
write
\begin{equation}
p^{\mu} = \overline{n}\cdot p \frac{n^{\mu}}{2} +(p_{\perp})^{\mu}
+n\cdot p \frac{\overline{n}^{\mu}}{2} = \mathcal{O} (\lambda^0) +
\mathcal{O} (\lambda^1) +\mathcal{O} (\lambda^2),
\end{equation}
since $p^+ p^- \sim p_{\perp}^2 \sim \lambda^2$. Therefore we have
three distinct energy scales $E$, $E\lambda$ and
$E\lambda^2$, making the effective theory  more complicated than the
HQET. It is similar to
the case of nonrelativistic QCD (NRQCD) for quarkonium states,
in which there are also three distinct scales $m$, $m\beta$ and
$m\beta^2$, where $m$ is the heavy quark mass and $\beta$ is the
typical velocity of a quark inside a quarkonium \cite{nrqcd}. 
The collinear quark can emit either a soft gluon with momentum
$k_s = E ( \lambda^2, \lambda^2, \lambda^2)$ or a collinear gluon with
$k_c = E(\lambda^2,1,\lambda)$ to the large momentum direction and
can still be on its mass shell. Due to the infrared sensitivity with
collinear loop momentum, the effective theory is more complicated, and
the relevant scales must be treated separately to obtain a consistent
power counting method. In the collinear effective theory, the power
counting in $1/E$ is troublesome, but the expansion in the small
parameter $\lambda$ offers a consistent power counting and there is no
mixing of operators with different powers of $\lambda$. This will be
discussed in detail in Section V.

The Lagrangian in the collinear effective theory can be obtained from
the full QCD Lagrangian at tree level by expanding it in powers of
$\lambda$. The full QCD Lagrangian for massless quarks and gluons is
given by
\begin{equation}
\mathcal{L}_{\mathrm{QCD}} = \overline{q} i\FMslash{D} q
-\frac{1}{4} G_{\mu\nu}^a G^{\mu\nu a},
\end{equation}
where the covariant derivative is $D_{\mu} = \partial_{\mu} +igT^a
A_{\mu}^a$, and $G_{\mu\nu}^a$ is the gluon field strength tensor. We
remove large momenta from the Lagrangian, similar to the method
employed in the HQET. The quark momentum is split as 
\begin{equation}
p=\tilde{p} +k, \ \ \tilde{p}\equiv (\overline{n}\cdot
p)\frac{n}{2}+p_{\perp}.
\end{equation}
The large part of the quark momentum $\overline{n} \cdot p$ and
$p_{\perp}$, denoted by $\tilde{p}$, will be removed by defining a new
field as
\begin{equation}
q (x) = \sum_{\tilde{p}} e^{-i\tilde{p} \cdot x} q_{n,p}(x).
\end{equation}
A label $p$ in $q_{n,p}$ refers to only the components
$\overline{n}\cdot p$ and $p_{\perp}$. The derivative $\partial_{\mu}$
on the field $q_{n,p}$ gives $O(\lambda^2)$ contributions. 

Now we introduce projection operators which project out large
components $\xi_{n,p}$ and small components $\xi_{\bar{n},p}$ in
the direction $n^{\mu}$ as 
\begin{equation}
\xi_{n,p} = \frac{\FMslash{n} \FMslash{\overline{n}}}{4}
q_{n,p}, \  \ \xi_{\bar{n},p} = \frac{\FMslash{\overline{n}}
\FMslash{n}}{4} q_{n,p}.
\end{equation}
The fields $\xi_{n,p}$, $\xi_{\bar{n},p}$ satisfy
\begin{equation}
\frac{\FMslash{n} \FMslash{\overline{n}}}{4} \xi_{n,p} = \xi_{n,p},\
 \frac{\FMslash{\overline{n}} \FMslash{n}}{4} \xi_{\bar{n},p} =
\xi_{\bar{n},p}, 
\end{equation}
and 
\begin{equation}
\FMslash{n} \xi_{n,p}=0, \ \FMslash{\overline{n}} \xi_{\bar{n},p}
=0.
\end{equation}
We can  eliminate the small component
$\xi_{\bar{n},p}$ at tree level by using the equation of motion
\begin{equation}
(\overline{n} \cdot p+\overline{n}\cdot iD) \xi_{\bar{n},p} =
(\FMslash{p}_{\perp} +i\FMslash{D}_{\perp}
)\frac{\FMslash{\overline{n}}}{2} \xi_{n,p},
\end{equation}
and the Lagrangian can be written in terms of $\xi_{n,p}$.
It is convenient to separate the collinear and soft parts in gluon
modes as $A^{\mu} = A_c^{\mu} +A_s^{\mu}$ in the covariant derivative
$D^{\mu}$, such that the covariant derivative involves only soft
gluons. The typical scale for the collinear gluons is $q^2 \sim
\lambda^2$, while the typical scale for the soft gluons is $k^2 \sim
\lambda^4$. Since the collinear gluon carries a large momentum
$\tilde{q} \equiv (\overline{n}\cdot q, q_{\perp})$, derivatives on
this field can yield order $\lambda^0$ and $\lambda^1$
contributions. To make this explicit, we extract the large momentum
part containing $\tilde{q}$  by redefining the field $A_c^{\mu} (x) = 
\sum_{\tilde{q}} e^{-i\tilde{q} \cdot x} A_{n,q}^{\mu} (x)$. Then the 
Lagrangian can be written as
\begin{eqnarray}
\mathcal{L}&=& \overline{\xi}_{n,p^{\prime}} \Bigl[ n\cdot iD -g
n\cdot A_{n,q} \nonumber \\
&&+ \Bigl(\FMslash{p}_{\perp} +i\FMslash{D}_{\perp} -
g\FMslash{A}_{n,q}^{\perp} \Bigr) \frac{1}{\overline{n} \cdot p
+\overline{n} \cdot iD -g \overline{n} \cdot A_{n,q}}
\Bigl(\FMslash{p}_{\perp} +i\FMslash{D}_{\perp} -g
\FMslash{A}_{n,q}^{\perp} \Bigr) \Bigr] \frac{\FMslash{\overline{n}}}{2}
\xi_{n,p}.
\label{lag1}
\end{eqnarray}
Here the summation over the labels $\tilde{p}$ and
$\tilde{p}^{\prime}$ and the phase factors for each collinear field
are suppressed. 
From now on, in order to simplify the notation further, we suppress
the label momenta for the collinear fields when
there can be no confusion. It should be understood that, when 
$\xi_n$ and $A_n^{\mu}$ appear, the summation on the label momentum
$p$, $q$, the large phases, and the conservation of the label momenta
are implied. The method to insert all the summations, the phases, and
the label momenta are nicely summarized in Ref.~\cite{bauer2}.

In order to
obtain the effective Lagrangian, we expand Eq.~(\ref{lag1}) in powers
of $\lambda$. In the power counting of the fields in $\lambda$, we
follow the procedure of moving all the dependence on $\lambda$ into
the interaction terms to make the kinetic terms of order
$\lambda^0$. This is done by assigning a $\lambda$ scaling to the
fields in the effective theory, as given in Table I \cite{bauer2}.

\begin{table}[t]
\caption{\label{lambdacount}Power counting for the effective theory
fields.} 
\begin{ruledtabular}
\begin{tabular}{ccccccc}
&heavy quark& collinear quark& soft gluon&&collinear gluon& \\ \hline
field& $h_v$& $\xi_{n,p}$& $A_s^{\mu}$ & $\overline{n}\cdot A_{n,q}$ &
$n\cdot A_{n,q}$& $A^{\perp}_{n,q}$ \\
scaling&$\lambda^3$ & $\lambda$ & $\lambda^2$ & $\lambda^0$
&$\lambda^2$ &$\lambda$ \\
\end{tabular}
\end{ruledtabular}
\end{table}

Bauer and Stewart \cite{bauer2} suggested a closed form to
include the effects of collinear gluons to all orders.
We define an operator $\overline{\mathcal{P}}$ 
which acts on products of effective theory fields. When acting on
collinear fields, $\overline{\mathcal{P}}$ gives the sum of large
momentum labels on fields minus the sum of large momentum labels on
conjugate fields. Then, for any function $f$, we have
\begin{eqnarray}
f(\overline{\mathcal{P}}) \Bigl( \phi_{q_1}^{\dagger} \cdots
\phi_{q_m}^{\dagger} \phi_{p_1} \cdots \phi_{p_n} \Bigr) &=&
f(\overline{n}\cdot p +\cdots +\overline{n} \cdot p_n -
\overline{n}\cdot q_1 -\cdots -\overline{n} \cdot q_m ) \nonumber \\
&&\times  \Bigl(
\phi_{q_1}^{\dagger} \cdots 
\phi_{q_m}^{\dagger} \phi_{p_1} \cdots \phi_{p_n} \Bigr).
\end{eqnarray}
The operator $\overline{\mathcal{P}}$ has mass dimension 1, but power
counting dimension $\lambda^0$. The conjugate operator
$\overline{\mathcal{P}}^{\dagger}$ acts only to its left and gives the
sum of large momenta on conjugate fields minus the sum of large
momenta on fields. 

Let us consider gauge symmetries of the effective theory. Since there
are several gluon modes, there are possible $SU(3)$ color gauge
transformations for each mode. We consider gauge symmetries that have
support over collinear momenta. The collinear effective theory is
invariant under a collinear nonabelian gauge transformation of the
form $U(x) = \exp [i\alpha^a (x) T^a]$. A set of these collinear gauge
transformations is a subset of all the gauge transformations, which
satisfies $\partial^{\mu} U \sim E (\lambda^2, 1,\lambda)$. It is
useful to decompose this collinear transformation into a sum over the
collinear momenta 
\begin{equation}
U(x) =\sum_Q e^{-iQ\cdot x} \mathcal{U}_Q,
\end{equation} 
where $\partial^{\mu} \mathcal{U}_Q \sim \lambda^2$.
When we expand the gauge transformation, we obtain simple
transformation rules for collinear fermions and gluons.
The transformation for collinear fermions and gluons are given by
\begin{equation}
\xi_n \rightarrow \mathcal{U} \xi_n, \ \ A_n^{\mu} \rightarrow
\mathcal{U} A_n^{\mu} \mathcal{U}^{\dagger} -\frac{1}{g} \mathcal{U}
\Bigl[ \Bigl( \overline{\mathcal{P}} \frac{n^{\mu}}{2}
+\mathcal{P}_{\perp}^{\mu}+(in\cdot \partial)
\frac{\overline{n}^{\mu}}{2}\Bigr) \mathcal{U}^{\dagger} \Bigr]. 
\label{cotran}
\end{equation}
Here $\mathcal{P}_{\perp}^{\mu}$
produces a sum of momenta of order $\lambda$, and the last term
produces a momentum of order $\lambda^2$. And the soft modes
transform as $A_s^{\mu} \rightarrow \mathcal{U} A_s^{\mu}
\mathcal{U}^{\dagger}$ under a collinear gauge transformation. 

Let us define a function $W$ of $\overline{n}\cdot A_n$ such that
$W^{\dagger} \xi_n$ is invariant under the transformation in
Eq.~(\ref{cotran}). The operators $W$ and $W^{\dagger}$ are defined as
\begin{equation}
W=\Bigl[ \exp \Bigl(\frac{1}{\overline{\mathcal{P}}} g\overline{n}
\cdot A_n \Bigr) \Bigr], \ \ W^{\dagger}=\Bigl[ \exp
\Bigl(g\overline{n} \cdot A_n^*
\frac{1}{\overline{\mathcal{P}}^{\dagger}}  \Bigr) \Bigr], 
\label{ws}
\end{equation}
which satisfy $W^{\dagger} W=1$. In the expansion of the exponential,
the $1/\overline{\mathcal{P}}$ acts to the right on all gluon fields
in the square bracket. Under a collinear gauge
transformation, $W$ transforms as \cite{bauer2} 
\begin{equation}
W\rightarrow \mathcal{U} W,
\end{equation}
which makes $W^{\dagger} \xi_n$ invariant under a collinear gauge
transformation. When we expand the exponential in $W$, we have an
infinite series of collinear gluons. But all of them are of order
$\lambda^0$, and should be included. 
The operator $\FMslash{\mathcal{P}}_{\perp}-g\FMslash{A}_n^{\perp}$ of
order $\lambda$ transforms as 
\begin{equation}
\FMslash{\mathcal{P}}_{\perp}-g\FMslash{A}_n^{\perp} \rightarrow
\mathcal{U} \Bigl(
\FMslash{\mathcal{P}}_{\perp}-g\FMslash{A}_n^{\perp} \Bigr)
\mathcal{U}^{\dagger}, 
\end{equation}
under a collinear gauge transformation.

\begin{figure}[b]
\begin{center}
\epsfig{file=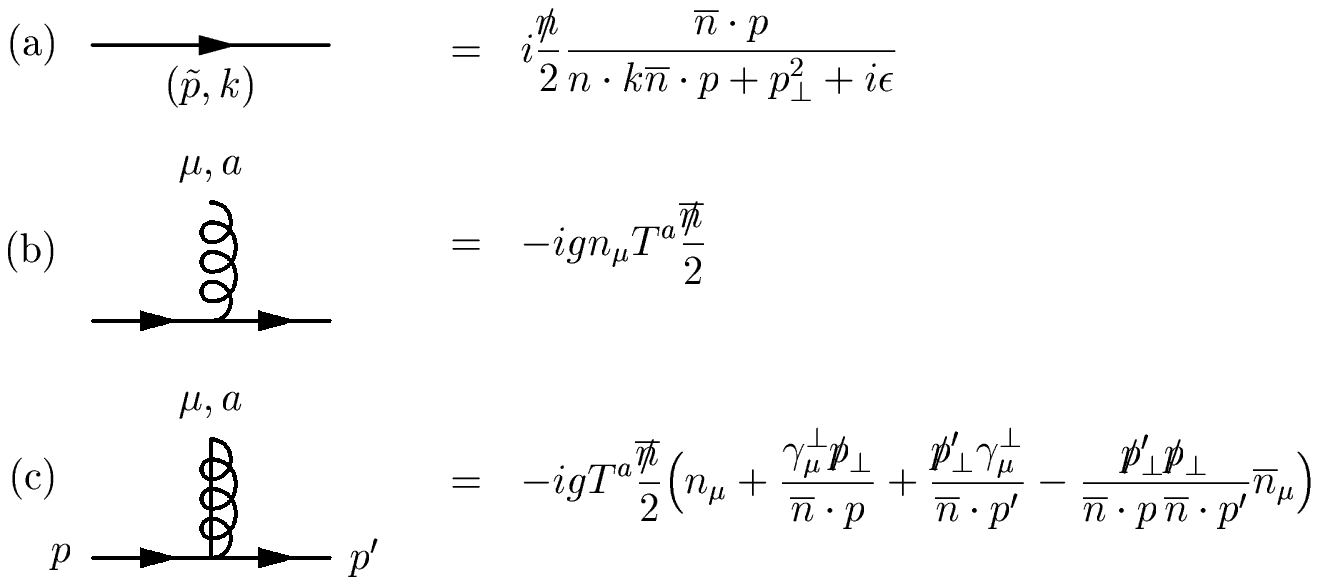}
\end{center}
\caption{Feynman rules for $\mathcal{L}_0$ to order $g$ in the
collinear effective theory: (a) collinear quark propagator with label
$\tilde{p}$ and residual momentum $k$, (b) collinear quark
interaction with one soft gluon, and (c) collinear quark interaction
with one collinear gluon, respectively.} 
\label{feynman}
\end{figure}

With these transformation properties, we can write the
Lagrangian $\mathcal{L}=\mathcal{L}_0+\mathcal{L}_1$ in a closed form
including an infinite number of collinear gluons as 
\begin{eqnarray}
\mathcal{L}_0 &=& \overline{\xi}_n \Bigl\{ n\cdot
(iD -g A_n) +(\FMslash{\mathcal{P}}_{\perp}
-g\FMslash{A}_n^{\perp} ) W \frac{1}{\overline{\mathcal{P}}}
W^{\dagger}
(\FMslash{\mathcal{P}}_{\perp}-g\FMslash{A}_n^{\perp}) 
\Bigr\} \frac{\FMslash{\overline{n}}}{2} \xi_n, \nonumber \\
\mathcal{L}_1 &=& \overline{\xi}_n \Bigl\{
i\FMslash{D}_{\perp} W \frac{1}{\overline{\mathcal{P}}} W^{\dagger}
(\FMslash{\mathcal{P}}_{\perp}-g\FMslash{A}_n^{\perp}) +
(\FMslash{\mathcal{P}}_{\perp} -g\FMslash{A}_n^{\perp} )W
\frac{1}{\overline{\mathcal{P}}} W^{\dagger}i\FMslash{D}_{\perp}
\Bigr\}  \frac{\FMslash{\overline{n}}}{2} \xi_n,
\label{eftlag}
\end{eqnarray}
where $\mathcal{L}_n \ (n=0,1)$ is the Lagrangian at order $\lambda^n$.  
The expression in Eq.~(\ref{eftlag}) is manifestly invariant under a
collinear gauge transformation, and we use the fact that for
any function $f$, $Wf(\overline{\mathcal{P}}) W^{\dagger} =
f(\overline{\mathcal{P}} -g \overline{n}\cdot A_n)$.

The Feynman rules for the propagator of a collinear quark and the
interaction vertices from $\mathcal{L}_0$ are shown in
Fig.~\ref{feynman}. Here $\gamma_{\perp}^{\mu}$ is defined as 
\begin{equation}
\gamma_{\perp}^{\mu} = \gamma^{\mu} -\frac{\FMslash{n}}{2}
\overline{n}^{\mu} -\frac{\FMslash{\overline{n}}}{2} n^{\mu}.
\end{equation}
There are other interaction vertices such as the one with two collinear
quark fields and two gluons, and those with triple gluons. We omit
them here since they do not contribute to one-loop corrections to
order $\lambda$ in dimensional regularization.

For a heavy quark, we 
employ HQET for the heavy quark field $h_v$. The effective
Lagrangian for HQET is given by
\begin{equation}
\mathcal{L}_{\mathrm{HQET}} = \overline{h}_v v\cdot iD h_v.
\label{hqet}
\end{equation}
The covariant derivative in Eq.~(\ref{hqet}) contains only soft
gluons because the heavy quark field does not couple to collinear 
gluons. According to the power counting in Table~\ref{lambdacount}, 
the corrections in $1/m_b$ in the HQET Lagrangian are suppressed by
$\lambda^2$ compared to the leading Lagrangian,
and we will not consider them here. 

\section{\label{repinvar}Reparameterization invariance}
When we decompose a quantity into a large part and a small part, the
decomposition is not unique. We can always shift the large part such
that a change in the small part compensates this change to make the
total quantity unchanged. The physics should be invariant under such a 
change. The invariance under this shift is called the
reparameterization invariance. In HQET, there is a reparameterization
invariance \cite{repar}. It means that the 
decomposition of the heavy quark momentum $p_b$ into $m_b v$ and the
residual momentum $k$ is not unique. Typically $k$ is of the order of
$\Lambda_{\mathrm{QCD}}$, which is much smaller than $m_b$. A small
change in the four velocity of the order of
$\Lambda_{\mathrm{QCD}}/m_b$ can be compensated by a change in the
residual momentum. The physics of heavy quarks should be invariant
under different decomposition of momenta. A consequence of this
reparameterization invariance is that the kinetic energy term in 
HQET is not renormalized to all orders. Besides, we can obtain
higher-dimensional operators for heavy-light currents using the
reparameterization invariance. And we can easily obtain the Wilson
coefficients and the anomalous dimensions of higher-dimensional
operators without any explicit calculation. 

A similar reparameterization invariance occurs in the collinear
effective theory. The energetic light quark momentum $p$ is given by
\begin{equation}
p^{\mu} = \frac{\overline{n} \cdot p}{2} n^{\mu} +p_{\perp}^{\mu}
+k^{\mu}.
\end{equation}
From now on, we will consider a small change of order $\lambda$,
neglecting terms of order $\lambda^2$, which can be included in 
a straightforward way. As in HQET, the decomposition of $p$ into 
$n$, $p_{\perp}$ is not unique. A small change in $n^{\mu}$ of order
$\lambda$ can be compensated by a change in $p_{\perp}^{\mu}$, 
\begin{equation}
n\rightarrow n+\frac{2\epsilon}{\overline{n}\cdot p}, \ \
p_{\perp}\rightarrow p_{\perp}-\epsilon,
\label{repar}
\end{equation}
where $\epsilon$ is of order $\lambda$. And the physics for collinear
quarks should be invariant under different decompositions of momenta.

Since $n$ satisfies $n^2=0$, the parameter $\epsilon$ must satisfy
$n\cdot \epsilon=0$, neglecting terms of order
$(\epsilon/\overline{n}\cdot p)^2$. The light quark spinor
$\xi_n$ must also change to preserve the constraint $\FMslash{n}
\xi_n =0$. Consequently, if $\xi_n$ changes as
$\xi_n \rightarrow \xi_n +\delta \xi_n$, $\delta \xi_n$ satisfies
\begin{equation}
\Bigl(\FMslash{n} +\frac{2\FMslash{\epsilon}}{\overline{n}\cdot p} \Bigr)
(\xi_n +\delta \xi_n) =0.
\end{equation}
To first order in $\epsilon/\overline{n}\cdot p$, one finds
\begin{equation}
\FMslash{n} \delta \xi_n =-\frac{2\FMslash{\epsilon}}{\overline{n}\cdot p}
\xi_n.
\label{rel}
\end{equation}
Therefore a suitable choice for the change in $\xi_n$ is
\begin{equation}
\delta \xi_n =-\frac{1}{\overline{n}\cdot p}
\frac{\FMslash{\overline{n}}}{2} \FMslash{\epsilon}\xi_n.
\label{delxi}
\end{equation}

The Lagrangian in Eq.~(\ref{eftlag}) must be invariant under the
combined changes
\begin{equation}
n\rightarrow n+\frac{2\epsilon}{\overline{n} \cdot p}, \ \ 
\xi_n\rightarrow e^{i\epsilon \cdot x} \Bigl(
1-\frac{1}{\overline{n} \cdot p}
\frac{\FMslash{\overline{n}}\FMslash{\epsilon}}{2}\Bigr) \xi_n,
\end{equation}
where the prefactor $e^{i\epsilon\cdot x}$ causes a shift $p_{\perp}
\rightarrow p_{\perp} -\epsilon$. In order to prove the
reparameterization invariance, it is convenient to write the
Lagrangian $\mathcal{L}$ as
\begin{equation}
\mathcal{L} = \overline{\xi}_n \Bigl\{ n\cdot
(iD +\mathcal{P}-g A_n) +(\FMslash{\mathcal{P}}_{\perp}
-g\FMslash{A}_n^{\perp} +i\FMslash{D}_{\perp} ) W
\frac{1}{\overline{\mathcal{P}}} W^{\dagger} 
(\FMslash{\mathcal{P}}_{\perp}-g\FMslash{A}_n^{\perp}
+i\FMslash{D}_{\perp})      
\Bigr\} \frac{\FMslash{\overline{n}}}{2} \xi_n, 
\label{replag}
\end{equation}
where we included $n\cdot \mathcal{P}$ which does not affect
the Lagrangian, but the addition makes the Lagrangian manifestly
invariant under a collinear gauge transformation.

The change of the Lagrangian is given by
\begin{eqnarray}
\delta \mathcal{L} &=& \overline{\xi}_n \Bigl[
\frac{2\epsilon}{\overline{n}\cdot p} \cdot
(\mathcal{P}_{\perp} -gA_{\perp} +iD_{\perp}) 
\nonumber \\
&&-\FMslash{\epsilon} W\frac{1}{\overline{\mathcal{P}}}W^{\dagger} 
(\FMslash{\mathcal{P}}_{\perp} -g\FMslash{A}_{\perp}
+i\FMslash{D}_{\perp}) -(\FMslash{\mathcal{P}}_{\perp}
-g\FMslash{A}_{\perp} +i\FMslash{D}_{\perp}) 
W\frac{1}{\overline{\mathcal{P}}}W^{\dagger}\FMslash{\epsilon} \Bigr]
\frac{\FMslash{\overline{n}}}{2} \xi_n.
\label{lchange}
\end{eqnarray}
The change $\delta \mathcal{L}$ vanishes, which can be easily seen
when we disregard gauge fields. Then the first line in
Eq.~(\ref{lchange}) exactly cancels the second line.
Therefore we have proved that the Lagrangian is
reparameterization invariant under a shift of order $\lambda$. As a
result, the kinetic energy terms appearing both in $\mathcal{L}_0$ and
$\mathcal{L}_1$ are not renormalized. The explicit 
calculation to show that the kinetic energy term at order
$\lambda$ is not renormalized at one loop is given in Appendix. 

We can make a stronger statement by combining the reparameterization
invariance and the collinear gauge invariance of the collinear
effective theory. In the Lagrangian $\mathcal{L}_1$ at order
$\lambda$, the kinetic energy part is given by  
\begin{equation}
\overline{\xi}_n \frac{2p_{\perp} \cdot
  i\partial_{\perp}}{\overline{n} \cdot p}
\frac{\FMslash{\overline{n}}}{2} \xi_n,
\label{firen}
\end{equation}
which is not renormalized due to the reparameterization invariance.
However, in order to make this part collinear gauge 
invariant, $\mathcal{P}_{\perp}$ should be replaced by
$\mathcal{P}_{\perp} -gA_{\perp}$. There is no
constraint from the collinear gauge invariance on whether
we should replace the derivative operator with a covariant derivative
including a soft gluon. However, if we require the invariance under 
ultrasoft gauge transformations \cite{bauer2}, the derivative operator
should be replaced by the covariant derivative.
Therefore the extension of the kinetic energy term which is invariant
under the collinear and the ultrasoft gauge transformation is given by 
\begin{equation}
\overline{\xi}_n \Bigl\{ (iD_{\perp})_{\mu} W
\frac{1}{\overline{\mathcal{P}}} 
W^{\dagger} (\mathcal{P}_{\perp}^{\mu} -gA_n^{\perp\mu}) +
(\mathcal{P}_{\perp}^{\mu} 
-gA_n^{\perp\mu})  W \frac{1}{\overline{\mathcal{P}}}
W^{\dagger}  ( iD_{\perp})_{\mu} \Bigr\}
\frac{\FMslash{\overline{n}}}{2} \xi_n.  
\end{equation}
This is not renormalized
to all orders in $\alpha_s$ due to the reparameterization invariance
and the gauge invariance. And the remaining part in $\mathcal{L}_1$ is 
not renormalized at one loop, hence the whole Lagrangian
$\mathcal{L}_1$ is not renormalized at leading logarithmic accuracy. 

We can fix the form of some corrections at order $\lambda$ from the
operators at $\lambda^0$ using the reparameterization invariance. For
example, the vector current $\overline{q}\gamma^{\mu} b$ in the full
theory is written as 
\begin{equation}
\overline{q}\gamma^{\mu} b \rightarrow \overline{\xi}_n
\Bigl(1+\frac{\overline{\FMslash{n}}}{2}
\frac{\FMslash{p}_{\perp}}{\overline{n} \cdot p} \Bigr) \gamma^{\mu}
h_v  = \overline{\xi}_n \gamma^{\mu} h_v +
\overline{\xi}_n \frac{\FMslash{\overline{n}}}{2}
\frac{\FMslash{p}_{\perp}}{\overline{n} \cdot p} \gamma^{\mu} h_v, 
\end{equation}
in the collinear effective theory to order $\lambda$. The collinear
gauge-invariant form of this operator is given by
\begin{equation}
\overline{\xi}_n \Bigl( 1 + \frac{\FMslash{\overline{n}}}{2}
(\FMslash{\mathcal{P}}_{\perp} 
-g\FMslash{A}_n^{\perp} ) W\frac{1}{\overline{\mathcal{P}}^{\dagger}}
\Bigr) \Gamma h_v,
\label{hlcur}
\end{equation}
where the second term is an operator for heavy-light currents at order
$\lambda$ in the effective theory.

\section{\label{match}Matching heavy-light currents}
We consider the matching of heavy-light currents of the form
$J=\overline{q} \Gamma b$, where $\Gamma$ denotes
$\gamma^{\mu}$ or $\gamma^{\mu}\gamma_5$. Below the scale
$\overline{n}\cdot p$, the hadronic current is matched onto
currents in the collinear effective theory and the HQET. This
introduces a new set of Wilson coefficients. 
We will match the current operators in the full theory with
the current operators in the collinear effective theory and the
HQET in a single step neglecting the sum of logarithms of
order $\ln (m_b/\overline{n}\cdot p)$, which is quite small since 
$m_b \sim \overline{n}\cdot p$. 

The vector-current operator $V^{\mu} = \overline{q} \gamma^{\mu} b$ in
the full theory can be matched to the 
effective theory as
\begin{equation}
V^{\mu} \rightarrow \sum_i C_i (\mu) J_i^{\mu}
+\sum_j B_j O_j^{\mu} + \sum_k A_k T_k^{\mu}.
\label{vectorc}
\end{equation}
The operators $J_i$ are the operators at leading order in $\lambda$,
and there are three such operators, which are given as
\begin{equation}
J_1^{\mu} =\overline{\xi}_n W \gamma^{\mu} h_v, \ \ J_2^{\mu}
=\overline{\xi}_n W v^{\mu} h_v, \ \  J_3^{\mu}
=\overline{\xi}_n W n^{\mu} h_v.
\label{leading}
\end{equation}

Similarly, $\{ O_j^{\mu} \}$ are a complete set of operators at order
$\lambda$. There are four such operators and a convenient basis for
these operators is given by 
\begin{eqnarray}
O_1^{\mu} &=& \overline{\xi}_n \frac{\FMslash{\overline{n}}}{2}
(\FMslash{\mathcal{P}}_{\perp} -g\FMslash{A}_{\perp}) W
\frac{1}{\overline{\mathcal{P}}^{\dagger}} \gamma^{\mu} h_v, \ 
O_2^{\mu} = \overline{\xi}_n \frac{\FMslash{\overline{n}}}{2}
(\FMslash{\mathcal{P}}_{\perp} -g\FMslash{A}_{\perp}) W
\frac{1}{\overline{\mathcal{P}}^{\dagger}} v^{\mu} h_v, \nonumber \\ 
O_3^{\mu} &=& \overline{\xi}_n \frac{\FMslash{\overline{n}}}{2}
(\FMslash{\mathcal{P}}_{\perp} -g\FMslash{A}_{\perp}) W
\frac{1}{\overline{\mathcal{P}}^{\dagger}} n^{\mu} h_v, \ 
O_4^{\mu} = \overline{\xi}_n
(\mathcal{P}_{\perp}^{\mu}-gA_{\perp}^{\mu}) W 
\frac{1}{\overline{\mathcal{P}}^{\dagger}} h_v.
\label{qlambda}
\end{eqnarray}
The operators in Eqs.~(\ref{leading}), and (\ref{qlambda}) are written
in such a way that they are manifestly invariant under a collinear
gauge transformation. 
We also include the nonlocal operators $T_k^{\mu}$ arising
from an insertion of the order $\lambda$ correction to the effective
Lagrangian into matrix elements of the leading-order currents, which
are defined as
\begin{equation}
T_k^{\mu} = i\int d^4 y T \Bigl\{ J_k^{\mu} (0),  \mathcal{L}_1 (y)  
\Bigr\}, \ (k=1,2,3).
\label{tfactor}
\end{equation}

Our goal is to calculate the Wilson coefficients $C_i(\mu)$,
$B_j(\mu)$ and $A_k(\mu)$ in the leading logarithmic
approximation. The Wilson coefficients are defined by requiring that
matrix elements of the vector current in the full 
theory are the same, to any order in $\lambda$, as matrix elements
calculated in the effective theory. 
Before we proceed to explicit calculation, note that there are
nontrivial relations between the coefficients $B_j (\mu)$ and
$C_j (\mu)$ imposed by the reparameterization invariance. This is
because operators of order $\lambda$
acting on a collinear quark field must always appear in certain
combinations with operators of order $\lambda^0$. 
In our case, there is a unique way in which the operators $O_i^{\mu}$
can be combined with $J_i^{\mu}$ in a reparameterization invariant
way, that is, 
\begin{eqnarray}
&&\langle \overline{\xi}_n \Bigl( 1 +\frac{\FMslash{\overline{n}}}{2}
\frac{\FMslash{p}_{\perp}}{\overline{n} \cdot p} \Bigr) \gamma^{\mu}
h_v\rangle +\cdots = \langle J_1^{\mu} \rangle +\langle O_1^{\mu}
\rangle +\cdots, \nonumber \\
&&\langle \overline{\xi}_n \Bigl( 1 +\frac{\FMslash{\overline{n}}}{2} 
\frac{\FMslash{p}_{\perp}}{\overline{n} \cdot p} \Bigr)v^{\mu} h_v
\rangle +\cdots = \langle J_2^{\mu} \rangle  + \langle O_2^{\mu}
\rangle +\cdots, \nonumber \\
&&\langle \overline{\xi}_n \Bigl( 1 +\frac{\FMslash{\overline{n}}}{2}
\frac{\FMslash{p}_{\perp}}{\overline{n} \cdot p} \Bigr) \Bigl(n^{\mu}
+\frac{2 p_{\perp}^{\mu}}{\overline{n} \cdot p} \Bigr) h_v \rangle 
+\cdots = \langle J_3^{\mu}\rangle  +\langle O_3^{\mu}\rangle +
2\langle O_4^{\mu} \rangle +\cdots.
\end{eqnarray}
This implies that, to all orders in perturbation theory,
\begin{equation}
B_i (\mu) = C_i (\mu), \ \ (i=1,2,3), \ \ B_4 (\mu) =2C_3 (\mu),
\end{equation}
and the coefficients $C_i (\mu)$ have
been calculated at leading logarithmic order in Ref.~\cite{bauer1}. 
This is our new result and it imposes an important constraint on the
theory, which must be obeyed by an explicit calculation.

The operator product expansion
of the axial vector current $A^{\mu}=\overline{q}\gamma^{\mu} \gamma_5
b$ can be simply obtained from
Eq.~(\ref{vectorc}) by replacing $\overline{q} \rightarrow
-\overline{q} \gamma_5$ if we perform the
calculation using the dimensional regularization with modified minimal
subtraction ($\overline{\mathrm{MS}}$) and the NDR scheme with
anticommuting $\gamma_5$. We can rewrite the axial current as $A^{\mu}
=-\overline{q} \gamma_5 \gamma^{\mu} b$. The $\gamma_5$ matrix acting
on the massless quark $q$ becomes $\pm 1$ depending on the chirality of
the quark. Chirality is conserved by the QCD interactions, so the
calculation of matching conditions proceeds just as in the vector
current case, except that $\overline{q}$ is replaced everywhere by
$\overline{q}\gamma_5$. At the end of the calculation, the $\gamma_5$
is moved back next to $h_v$, producing a compensating minus sign for
$\gamma^{\mu} \gamma_5$, but neither for $v^{\mu} \gamma_5$ nor for
$n^{\mu} \gamma_5$. Thus, for axial vector currents, all the
coefficients are the same in magnitude, and only $C_1$, $B_1$, and
$A_1$ do not change sign, while all the remaining coefficients change
sign.

Bauer et al. \cite{luke,bauer1} have explicitly showed that the
collinear effective theory, indeed, reproduces the infrared behavior
of the full theory by including the effects of collinear gluons. 
Once we know that the effective theory reproduces the long-distance
physics of the full theory, the matching procedure is independent of
any long-distance physics such as infrared singularities,
nonperturbative effects and the choice of external states. Thus there
is a freedom in choosing the external states and the infrared
regularization scheme. We find it most convenient to perform the
matching of QCD onto the collinear and the heavy quark effective
theory using on-shell external quark states and dimensional
regularization for both the ultraviolet and infrared divergences
encountered in calculating loop diagrams. This scheme has the great
advantage that all loop diagrams in the effective theory vanish, since
there is no mass scale other than the renormalization scale $\mu$. It
means that matrix elements in the effective theory are given by their
tree-level expressions. We assign momentum such that the incoming
heavy quark has momentum $p_b = m_b v+k$ (with $2v\cdot k
+k^2/m_b=0$), while the outgoing light energetic quark carries
momentum $p = En+ p_{\perp}+k^{\prime}$ (with $2En\cdot k^{\prime}
+p_{\perp}^2 =0$). 

The matrix elements of operators can be written as
\begin{equation}
\langle J_1^{\mu} \rangle = \overline{u}_e (n,s) \gamma^{\mu} u_h
(v,s_b), \
\langle O_1^{\mu} \rangle = \overline{u}_e (n,s)
\frac{\FMslash{\overline{n}}}{2}
\frac{\FMslash{p}_{\perp}}{\overline{n} \cdot p} \gamma^{\mu} u_h
(v,s_b),
\end{equation}
where $u_e (n,s)$ and $u_h (v,s_b)$ are on-shell spinors for a
massless, energetic quark field $\xi_n$ in the collinear effective
theory, and a heavy quark field $h_v$ in the HQET, respectively. They
satisfy $\FMslash{n} u_e (n,s) =0$ and $\FMslash{v} u_h (v,s_b) = 
u_h (v,s_b)$. We compute, in the full theory, the vector current
matrix element between on-shell quark states at one-loop order in
order to do the matching. The relations of the
heavy quark spinors and the light quark spinors between QCD and the
effective theory are given by
\begin{eqnarray}
u_b (p_b, s_b) &=& \Bigl (1+\frac{\FMslash{k}}{2m_b} \Bigr) u_h
(v,s_b) +O(1/m_b^2), \nonumber \\
u_q (p,s) &=& \Bigl(1-\frac{\FMslash{\overline{n}}}{2}
\frac{\FMslash{p}_{\perp}}{\overline{n} \cdot p} \Bigr) u_e (n,s) +
O(\lambda^2).
\end{eqnarray}
The correction to the heavy quark field, which involves $\FMslash{k}$,
is suppressed by $\lambda^2$, and it is discarded in our matching at
order $\lambda$. 

We match the coefficients at one loop by employing the dimensional
regularization in $D=4-2 \epsilon$ dimensions. In the full theory,
there is no ultraviolet divergence due to current conservation. The
residue at the physical mass pole in the propagator is infrared in
nature, and it should be added to the vertex correction. 
The residue at the physical mass pole for the heavy quark in the
$\overline{\mathrm{MS}}$ scheme at order $\alpha_s$ is given by
\cite{manohar} 
\begin{equation}
R^{(1)}_b = -\frac{\alpha_s C_F}{4\pi} \Bigl( \frac{2}{\epsilon} +4
-6\ln \frac{m_b}{\mu} \Bigr), 
\end{equation}
and in the HQET, the residue at order $\alpha_s$ is given as
\begin{equation}
R^{(1)}_h = -\frac{\alpha_s C_F}{4\pi} \frac{2}{\epsilon}.
\end{equation}
The residue for the light quark 
at order $\alpha_s$ in the collinear effective theory is the same as
the residue in the full theory, and it is given as
\begin{equation}
R^{(1)}_q = R^{(1)}_{\xi_n} = \frac{\alpha_s C_F}{4\pi}
\frac{1}{\epsilon}.  
\end{equation}
Since the residues for the light quarks are the same, they cancel each
other when we match both theories.

The matrix element of the vector
current between free quark states with the residues of the external
quarks in the full theory can be expressed
in terms of the matrix elements in the 
collinear and the heavy quark effective theory as
\begin{eqnarray}
\langle \overline{q} \gamma^{\mu} b \rangle &=&  \Bigl\{
1-\frac{\alpha_s C_F}{4\pi} \Bigl[ \frac{1}{\epsilon^2}
+\frac{5}{2\epsilon} -\frac{2}{\epsilon} \ln \frac{xm_b}{\mu}
\nonumber \\
&&+2\ln^2 \frac{xm_b}{\mu}  +\frac{3x-2}{1-x} \ln x +Li_2 (1-x)
+\frac{\pi^2}{12} +6 \Bigr] \Bigr\} \langle J_1^{\mu}
+O_1^{\mu}\rangle \nonumber \\
&+& \frac{\alpha_s C_F}{4\pi} \Bigl[ \frac{2}{1-x} +\frac{2x}{(1-x)^2}
\ln x \Bigr] \langle J_2^{\mu} +O_2^{\mu}
\rangle \nonumber \\
&+& \frac{\alpha_s C_F}{4\pi} \Bigl[ -\frac{x}{1-x}
+\frac{x(1-2x)}{(1-x)^2} \ln x \Bigr] \langle J_3^{\mu}
+O_3^{\mu} +2 O_4^{\mu} \rangle,
\label{matching}
\end{eqnarray}
where $x=\overline{n}\cdot p/m_b = 2E/m_b$ and $Li_2 (x)$ is the
dilogarithmic function. Here we have confirmed the consequence of the 
reparameterization invariance at one loop explicitly. The infrared
behavior of the full QCD is reproduced in the collinear effective
theory, and the infrared divergences in both theories cancel in
matching.   

The Wilson coefficients $C_i$ for $J_i^{\mu}$ 
at the renormalization scale $\mu$ are given by
\begin{eqnarray}
C_1 (\mu) &=& 1-\frac{\alpha_s C_F}{4\pi} \Bigl[ 2 \ln^2
\Bigl(\frac{xm_b}{\mu}\Bigr) -5 \ln \frac{m_b}{\mu} + \frac{3x-2}{1-x}
\ln x +2 Li_2 (1-x) +\frac{\pi^2}{12}+6 \Bigr] , \nonumber \\
C_2 (\mu) &=& \frac{\alpha_s C_F}{4\pi} \Bigl[ \frac{2}{1-x} 
+\frac{2x}{(1-x)^2} \ln x \Bigr], \nonumber \\
C_3 (\mu) &=&  \frac{\alpha_s C_F}{4\pi} \Bigl[ -\frac{x}{1-x}
+\frac{x(1-2 x)}{(1-x)^2} \ln x \Bigr],
\label{wilson}
\end{eqnarray}
and the coefficients $B_j$ are given as
\begin{equation}
B_i (\mu) = C_i  (\mu) \ (i=1,2,3), \ \ B_4 (\mu) = 2 C_3 (\mu). 
\label{bis}
\end{equation}
This relation is expected from the reparameterization invariance, and
the operators $O_i^{\mu}$ have the 
same anomalous dimension as those of the leading operators
$J_i^{\mu}$. The explicit calculation that the operators $O_1$ to
$O_4$ have the same ultraviolet behavior as their corresponding
leading-operators is shown in Section~\ref{rg}.

The coefficients $A_i$ are given by the
product of those for $J_i^{\mu}$ and $\mathcal{L}_1$, and they are
given by
\begin{equation}
A_i (\mu) = C_i (\mu).
\label{ai}
\end{equation}
The fact that the Wilson coefficients $A_i$ are the same as $C_i$ is
because the effective Lagrangian $\mathcal{L}_1$ at order $\lambda$ is 
not renormalized at leading logarithmic order.

\section{\label{rg}Renormalization group improvement}
The perturbative expansion of the Wilson coefficients contains large
logarithms of the type $[\alpha_s \ln (2E/\mu)]^n$, which should be
summed to all orders. We employ the renormalization group to improve
one-loop results. The reason why we choose $\lambda$ as the small
parameter is because various operators with different orders of
$\lambda$ do not mix in this power counting. If we choose to
expand in powers of $1/E$, when we renormalize operators, 
a factor $E$ in the numerator could be induced from loop
calculations. This is expected since the propagator of a collinear
quark explicitly involves $E$ in the $1/E$ expansion. Therefore
higher-dimensional operators in $1/E$ can mix with 
those operators with one less power of $m_b$ or $E$, and a power
counting in $1/E$ is inappropriate. However, if we
expand the effective Lagrangian in powers of $\lambda$, such mixing
never occurs, and we can do the power counting in $\lambda$
consistently.

In general, the
coefficients of the operators with the same power of $\lambda$ mix
into themselves and satisfy a renormalization group equation of the
form
\begin{equation}
\mu \frac{d}{d\mu} C(\mu) =\gamma(\mu) C(\mu).
\label{rgeq}
\end{equation}
Since Eq.~(\ref{rgeq}) is homogeneous, we can reproduce the
exponentiation of Sudakov logarithm.

The renormalization of the operators $J_i^{\mu}$ at order $\lambda^0$
was performed in Ref.~\cite{bauer1}. The counterterm for the
operators $J_i^{\mu}$ in the effective theory using the Feynman gauge
is given by
\begin{equation}
Z_i = 1+\frac{\alpha_s C_F}{4\pi} \Bigl[ \frac{1}{\epsilon^2}
-\frac{2}{\epsilon} \ln \frac{2E}{\mu} +\frac{5}{2\epsilon} \Bigr].
\end{equation}
This counterterm is the same for all $J_i^{\mu}$, and is independent
of the Dirac structure of the operators since the propagators and the
vertices in the collinear effective theory do not alter the Dirac
structure of the operators. Furthermore there is no operator
mixing. The anomalous dimensions are given by
\begin{equation}
\gamma_i = Z_i^{-1} \Bigl( \mu \frac{\partial}{\partial \mu} +\beta
\frac{\partial}{\partial g} \Bigr) Z_i,
\end{equation}
where 
\begin{equation}
\mu \frac{\partial}{\partial \mu} Z_i = \frac{\alpha_s
(\mu)C_F}{2\pi \epsilon}, \ \
\beta \frac{\partial}{\partial g} Z_i = -\frac{\alpha_s C_F}{2\pi}
\Bigl( \frac{1}{\epsilon} -2\ln \frac{2E}{\mu} +\frac{5}{2} \Bigr).
\label{anomd}
\end{equation}
Here we have used $\beta =-g\epsilon +O(g^3)$. This gives the
anomalous dimension
\begin{equation}
\gamma_i = -\frac{\alpha_s (\mu) C_F}{2\pi} \Bigl( \frac{5}{2}-2\ln
\frac{2E}{\mu} \Bigr).
\end{equation}
The divergence in Eq.~(\ref{anomd}) is cancelled, and solving the
renormalization group equation Eq.~(\ref{rgeq}), we
obtain 
\begin{equation}
C_i (\mu) = \Bigl( \frac{\alpha_s (\mu)}{\alpha_s (2E)}
\Bigr)^{(C_F/2\beta_0) (5-8\pi/\beta_0 \alpha_s)}
\Bigl(\frac{2E}{\mu} \Bigr)^{2C_F/\beta_0} C_i (2E),
\end{equation}
where $\beta_0 = 11-2n_f/3$, and $C_i (2E)$ are the Wilson
coefficients at $\mu=\overline{n} \cdot p =2E$, as given in
Eq.~(\ref{wilson}).

\begin{figure}[t]
\begin{center}
\vspace{1.0cm}
\epsfig{file=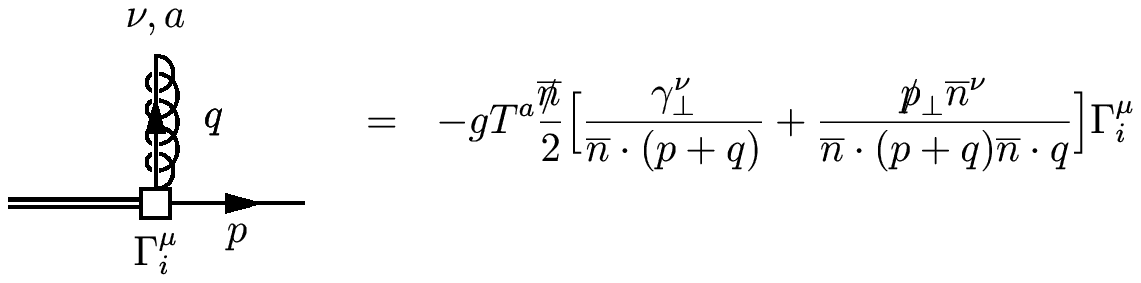}
\end{center}
\vspace{-1.0cm}
\caption{Feynman rules for the operator $O_i^{\mu}$ ($i=1,2,3$)
  containing a collinear gluon at
order $\lambda$. Here $\Gamma_i^{\mu}=\gamma^{\mu}$, $v^{\mu}$ and
$n^{\mu}$ for $i=1,2,3$ respectively. The momentum of the gluon is
outgoing.} 
\label{feyn1}
\end{figure}

\begin{figure}[t]
\begin{center}
\epsfig{file=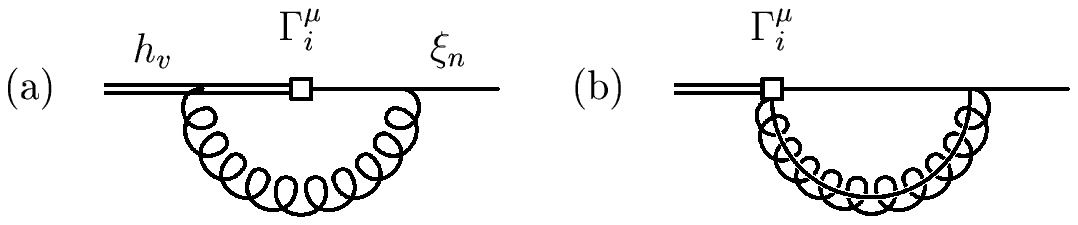}
\end{center}
\caption{Feynman diagrams for the renormalization of $O_i^{\mu}$
($i=1,2,3$) at one loop.}
\label{renlam1}
\end{figure}

At order $\lambda$, we need to renormalize the operators
$O_i^{\mu}$. Let us first consider the renormalization of $O_1^{\mu}$
to $O_3^{\mu}$. The
Feynman rules for the vertex from these operators with a
collinear gluon are given in Fig.~\ref{feyn1}. 
The Feynman diagrams to renormalize the operators $O^{\mu}_i$
($i=1,2,3$) at order $\alpha_s$ are shown in  
Fig.~\ref{renlam1}. Since the loop calculation does not alter the Dirac
structure, we can treat the renormalization of these operators in the
same way for all the three operators. The Feynman diagrams in
Fig.~\ref{renlam1} give the amplitude   
\begin{equation}
M^{(1)\mu}_i= -\frac{\alpha_s C_F}{4\pi} O_i^{\mu}
\Bigl[ \frac{1}{\epsilon^2} +(2-2\ln \frac{\overline{n}\cdot p}{\mu})
\frac{1}{\epsilon} \Bigr].
\end{equation}
Note that there is no mixing for the operators $O_i^{\mu}$. If we add
the residues from the propagators of a heavy quark and a 
collinear quark, we have the counterterm
\begin{equation} 
Z^{(1)}_i = 1+\frac{\alpha_s C_F}{4\pi} \Bigl[ \frac{1}{\epsilon^2}
-\frac{2}{\epsilon} \ln \frac{2E}{\mu} +\frac{5}{2\epsilon} \Bigr], \
\ (i=1,2,3),
\label{counter}
\end{equation}
which is identical to the counterterm for the leading operators
$J^{\mu}_i$. We can do the same calculation for the operator
$O_4^{\mu}$ and it turns out that the operator $O_4^{\mu}$ has the
same dependence on $\epsilon$ as $J_3^{\mu}$. And the counterterm is
also given by Eq.~(\ref{counter}).
Therefore the operators $O_i^{\mu}$ ($i=1,\cdots, 4$) have the same
anomalous dimensions as the leading operators. This is the explicit
proof of the reparameterization invariance at one loop and order
$\lambda$. 

For the time-ordered products $T_k^{\mu}$, the anomalous dimensions
are the same as those of $J_k^{\mu}$ because the
Lagrangian $\mathcal{L}_1$ in defining $T_k^{\mu}$ is not
renormalized at one loop. Here we see that the reparameterization
invariance and the gauge invariance influence the structure
of the theory.  Furthermore, since there is no mixing,
the perturbative corrections to heavy-light currents take a simple
form to order $\lambda$.

\section{\label{form}Application to form factors}
As an application of the collinear effective
theory, we can consider the form factors for $B$ mesons into light
mesons. We consider the kinematic region in which the energy $E$ of
the light quark is large,
\begin{equation}
E =\frac{m_b^2 -q^2}{2m_b} \sim \frac{m_b}{2},\ \ q=p_b -p_q,
\end{equation}
which equivalently means that the momentum transfer squared through
the weak current is small $q^2 \ll m_b^2$. In this case, the
off-shellness of the light quark is $p_q^2 =2Ek_+$, where $k_+ \sim
\Lambda_{\mathrm{QCD}}$, thus $\lambda \sim
\sqrt{\Lambda_{\mathrm{QCD}}/m_b}$. Therefore our formulation to order
$\lambda$ gives the correction to the form factors at order
$\sqrt{\Lambda_{\mathrm{QCD}}/m_b}$. For simplicity, we will consider
the form factors for the vector and the axial vector currents.

The form factors for $\overline{B}$ decays into light pseudoscalar and
vector mesons from the vector current $V^{\mu} =\overline{q}
\gamma^{\mu} b$, and the axial vector current $A^{\mu} = \overline{q}
\gamma^{\mu} \gamma_5 b$ are defined as
\begin{eqnarray}
\langle P (p^{\prime})| V^{\mu} |\overline{B}
(p)\rangle &=& f_+ (q^2) \Bigl[ p^{\mu} + p^{\prime \mu} -\frac{M^2
-m_P^2}{q^2} q^{\mu}\Bigr] + f_0 (q^2) \frac{M^2 -m_P^2}{q^2} q^{\mu},
\nonumber \\
\langle V (p^{\prime}, \epsilon^*)| V^{\mu} |\overline{B} (p) \rangle
&=& \frac{2V(q^2)}{M+m_V} i\epsilon^{\mu\nu \alpha \beta}
\epsilon^*_{\nu} p^{\prime}_{\alpha} p_{\beta}, \nonumber \\
\langle V (p^{\prime}, \epsilon^*)| A^{\mu} |\overline{B} (p) \rangle
&=& 2m_V A_0 (q^2) \frac{\epsilon^* \cdot q}{q^2} q^{\mu} +(M+m_V) A_1
(q^2) \Bigl[ \epsilon^{*\mu} -\frac{\epsilon^* \cdot q}{q^2} q^{\mu}
\Bigr] \nonumber \\
&&-A_2 (q^2) \frac{\epsilon^* \cdot q}{M+m_V} \Bigl[p^{\mu} +p^{\prime
\mu} -\frac{M^2-m_V^2}{q^2} q^{\mu} \Bigr],
\label{parform}
\end{eqnarray}
where $q= p-p^{\prime}$, $m_P$ ($m_V$) is the mass of the pseudoscalar
(vector) meson, $\epsilon^*_{\mu}$ is the polarization vector of the
vector meson, and $M$ is the mass of a $\overline{B}$ meson. We use
the sign convention $\epsilon^{0123} =-1$. 

We can calculate these form factors systematically in powers of
$\lambda$ in the collinear effective theory. The
matrix elements in the full theory are matched to the matrix elements
in the collinear effective theory using Eq.~(\ref{vectorc}).
However, here we do not include interactions where a
collinear gluon is exchanged with the spectator quarks inside a $B$
meson. In Ref.~\cite{beneke} it was argued that these spectator
effects could be of the same order in $\lambda$ and $1/m_b$ as the
soft contributions, but they are suppressed by a power of $\alpha_s
(\sqrt{m_b \Lambda_{\mathrm{QCD}}})$. They are therefore just as
important as the one-loop corrections to the matching coefficients
such as $C_i (\mu)$. Here we apply the collinear effective theory to
the soft contributions only. It means that the effective theory
applies to light mesons produced in an asymmetric configuration, in
which a single quark from the $b$ decay carries almost all the
momentum. 

If we consider this process as light-cone dominated, this is
not a typical configuration. A typical configuration is for both quark
and antiquark have nearly equal momentum. And spectator interactions
can play an important role in this configuration. 
In heavy-to-heavy transitions such as $B\rightarrow D$ in
the heavy quark limit, the interactions of a heavy quark with the soft
degrees of freedom around the heavy quark do not change even when
there is a transition. On the contrary, in
heavy-to-light transitions, the
soft degrees of freedom around the heavy quark experience an abrupt
change. If an energetic quark and the soft
degrees of freedom move somehow elastically with almost the same
velocity, we can safely consider the interaction of an energetic quark
with the soft degrees of freedom in terms of the collinear effective
theory. This corresponds to the soft contribution to form factors. If
only an energetic quark is pushed to the light-cone direction, the
soft degrees of freedom around the heavy quark should arrange
themselves to follow the energetic quark to form light mesons. In this
process, hard gluons should be exchanged between the energetic quark
and the previous soft degrees of freedom in the heavy quark. This
corresponds to the hard spectator interaction. This hard spectator
interaction should be considered separately, and we leave the hard
spectator contributions for future study.

A convenient way to evaluate hadronic matrix elements in the effective
theory is to associate the spin wave function
\begin{equation}
\mathcal{M} (v) = \sqrt{M} \frac{1+\FMslash{v}}{2} \left(
\begin{array}{c} -\gamma_5, \\
\FMslash{\epsilon},
                                              \end{array}
\right) \ \ \begin{array}{c}
\mbox{pseudoscalar meson } \ P,\\
\mbox{vector meson } \ V,
            \end{array}
\end{equation}
with the eigenstates of the effective Lagrangian, where $M$ is the
mass of the meson. The form factors in the effective theory can be
written as 
\begin{equation}
\langle L (n) | \overline{\xi}_n \Gamma h_v |\overline{B} (v)\rangle =
\mathrm{tr} \Bigl[ A_L (E) \overline{\mathcal{M}}_L \Gamma
\mathcal{M}_B \Bigr], \ \ (L=P, V)
\end{equation}
where $\Gamma$ denotes a Dirac structure, and 
\begin{eqnarray}
\overline{\mathcal{M}}_L &=&
  \left( \begin{array}{c}
-\gamma_5, \\
\FMslash{\epsilon}^*,
                                              \end{array}
\right) \frac{\FMslash{\overline{n}} \FMslash{n}}{4} \ \
  \begin{array}{c} 
L= P,\\
L=V,
            \end{array} \nonumber \\
\mathcal{M}_B &=& \frac{1+\FMslash{v}}{2} (-\gamma_5)
\end{eqnarray}
are the spin wave functions associated with a light meson and a
$\overline{B}$ meson respectively. The normalization factor $\sqrt{M}$
appearing in $\mathcal{M}$ is absorbed in $A_L(E)$. The function
$A_L (E)$ contains the long-distance dynamics, and it is independent
of the Dirac structure $\Gamma$ in the current. The most general form
for $A_L(E)$ is given by
\begin{equation}
\Xi_L (E) = \xi_{1L}(E) + \xi_{2L} (E) \FMslash{v} + \xi_{3L} (E)
\FMslash{n} +\xi_{4L} \FMslash{n}\FMslash{v},
\end{equation}
but due to the properties of the projection operators in
$\overline{\mathcal{M}}_L$ and $\mathcal{M}_B$, not all of them are
independent. For $L=P$, there is one independent term, and for $L=V$,
there are two independent terms. 

Charles et al. \cite{charles} have shown that there are only
three independent matrix elements in heavy-to-light transitions 
by employing the HQET and the large-energy effective theory to obtain
the leading result in $1/E$. However this is not sufficient to
describe heavy-to-light decays because interactions with collinear
gluons should be included. Though the argument is different,
there are also three independent matrix
elements in the collinear effective theory.

At order $\lambda$, we have the form factors of the form
\begin{equation} 
\langle L(n) | \overline{\xi}_n p_{\perp}^{\mu} \Gamma h_v
|\overline{B} (v)\rangle = \mathrm{tr} \ \Bigl[ A_L^{\mu} (E)
\overline{\mathcal{M}}_L \Gamma \mathcal{M}_B \Bigr], 
\end{equation}
where $A_L^{\mu} (E)$ contains the long-distance dynamics and they are
independent of the Dirac structure $\Gamma$ of the current. Since the
operator is 
proportional to $p_{\perp}^{\mu}$, the only allowed vector component
for $A_L^{\mu}$ is $\gamma_{\perp}^{\mu}$. Therefore the most general
form for $A_L^{\mu}$ is given by
\begin{equation}
A_L^{\mu} (E) = \gamma^{\mu}_{\perp} \Bigl[ a_{1L}(E) + a_{2L} (E)
\FMslash{v} + a_{3L} (E) \FMslash{n} +a_{4L}
\FMslash{n}\FMslash{v}\Bigr].  
\end{equation}
As in the case of $\Xi_L (E)$, all the terms are not independent due to
the projection operators in $\overline{\mathcal{M}}_L$, and
$\mathcal{M}_B$. For $L=P$, there is only one independent term, and
for $L=V$, there are two independent terms. Similarly, the matrix
elements of the time-ordered products $T_i$ can be written as
\begin{equation}
\langle L (n) | i\int d^4 y \ T\Bigl\{ J_i^{\mu} (0), \mathcal{L}_1 (y)
\Bigr\} |\overline{B}(v)\rangle =  \mathrm{tr} \ \Bigl[ B_L^{\mu} (E)
\overline{\mathcal{M}}_L \Gamma \mathcal{M}_B \Bigr], 
\end{equation}
and the most general form for $B_L^{\mu} (E)$ is written as
\begin{equation}
B_L^{\mu} (E) = \gamma_{\perp}^{\mu} \Bigl[ b_{1L}(E) + b_{2L} (E)
\FMslash{v} + b_{3L} (E) \FMslash{n} +b_{4L}
\FMslash{n}\FMslash{v}\Bigr],
\end{equation}
because $\mathcal{L}_1$ is of order $\lambda$ and it typically depends
on $p_{\perp}^{\mu}$. Here also we have one independent term for
$L=P$, and two independent terms for $L=V$.  

In summary, we can write the parameters describing the
long-distance physics as 
\begin{eqnarray}
\Xi_P (E) &=& 2E \xi_P, \ \Xi_V (E)= E\FMslash{n} \Bigl( \xi_{\perp}
-\frac{\FMslash{v}}{2} \xi_{\parallel} \Bigl), \nonumber \\
A_P^{\mu} (E) &=& \frac{a_P}{2} \gamma^{\mu}_{\perp}, \ A_V^{\mu} (E)
= \gamma^{\mu}_{\perp} \frac{\FMslash{n}}{2} \Bigl( 
a_{V1} +\frac{\FMslash{v}}{2} a_{V2} \Bigr), \nonumber \\
B_P^{\mu} (E) &=& b_P \gamma_{\perp}^{\mu}, \ B_V^{\mu} (E) =
\gamma^{\mu}_{\perp} \frac{\FMslash{n}}{2} \Bigl( b_{V1}
-\frac{\FMslash{v}}{2} b_{V2} \Bigr).
\end{eqnarray}
Note that the convention for the longitudinal form factor
$\xi_{\parallel}$ is the same as that of Ref.~\cite{beneke}, and is
related to the corresponding form factor $\zeta_{\parallel}$ defined
in Ref.~\cite{charles} by $\xi_{\parallel} (E) = (m_V/M)
\zeta_{\parallel} (E)$.  
The matrix elements of all the operators can expressed in terms of
these nonperturbative parameters. At order $\lambda^0$, the matrix
elements for pseudoscalar bosons are given by
\begin{eqnarray}
\langle P |\overline{\xi}_n \gamma^{\mu} h_v |\overline{B}\rangle &=&
2E \, \xi_P n^{\mu}, \ \langle P |\overline{\xi}_n
\gamma^{\mu}\gamma_5 h_v |\overline{B}\rangle =0, \nonumber \\
\langle P|\overline{\xi}_n v^{\mu} h_v |\overline{B}\rangle &=& 2E \, 
\xi_P v^{\mu}, \ \langle P|\overline{\xi}_n v^{\mu} \gamma_5  h_v
|\overline{B}\rangle =0, \nonumber \\
 \langle P|\overline{\xi}_n n^{\mu} h_v |\overline{B}\rangle &=&
2E \, \xi_P n^{\mu}, \ \langle P|\overline{\xi}_n n^{\mu} \gamma_5 h_v 
|\overline{B}\rangle =0.
\end{eqnarray}
For vector mesons, the matrix elements are written as
\begin{eqnarray}
\langle V | \overline{\xi}_n \gamma^{\mu} h_v|\overline{B}\rangle &=&
2E \, \xi_{\perp} i\epsilon^{\mu \nu \alpha \beta} \epsilon^*_{\nu}
n_{\alpha} v_{\beta}, \nonumber \\
\langle V | \overline{\xi}_n v^{\mu} h_v|\overline{B}\rangle &=&
\langle V | \overline{\xi}_n n^{\mu} h_v|\overline{B}\rangle =0,
\nonumber \\ 
\langle V | \overline{\xi}_n \gamma^{\mu} \gamma_5
h_v|\overline{B}\rangle &=& 2E \, \xi_{\perp} \Bigl(\epsilon^{*\mu}
-(\epsilon^* \cdot v) n^{\mu} \Bigr) 
+2E \, \xi_{\parallel} (\epsilon^* \cdot v) n^{\mu}, \nonumber \\
\langle V | \overline{\xi}_n v^{\mu}\gamma_5 h_v|\overline{B}\rangle
&=& -2E \, \xi_{\parallel} (\epsilon^* \cdot v ) v^{\mu}, \nonumber \\ 
\langle V | \overline{\xi}_n n^{\mu}\gamma_5 h_v|\overline{B}\rangle
&=& -2E \, \xi_{\parallel} (\epsilon^* \cdot v ) n^{\mu}. 
\end{eqnarray}

Using the above relations, we can determine the heavy-to-light form
factors at leading order in $\lambda$ and $\alpha_s$.
\begin{eqnarray}
f_+ (q^2) &=&  \frac{f_0 (q^2)}{X} = \xi_P (E), \ \ 
\frac{2\hat{m}_V}{X} A_0 (q^2) = \xi_{\parallel} (E), \nonumber \\
\frac{1+\hat{m}_V}{X} A_1 (q^2) &=& \frac{V (q^2)}{1+\hat{m}_V} =
\xi_{\perp} (E), \ \ 
\frac{A_2(q^2)}{1+\hat{m}_V} = \xi_{\perp} (E)-\xi_{\parallel},
\end{eqnarray}
where $X=2E/M$, $\hat{m_V} = m_V/M$. From the results in
Section IV, we can include the perturbative corrections, which change
the relation between form factors. We find that, at leading order in
$\lambda$ and at leading logarithmic order in $\alpha_s$, 
\begin{eqnarray}
f_+ &=& \xi_P (E) \Bigl[ C_1 +\frac{X}{2} C_2 + C_3
\Bigr], \ \ \frac{f_0}{X} = \xi_P (E) \Bigl[ C_1 +\Bigl(1-\frac{X}{2} 
\Bigr) C_2 +C_3\Bigr], \nonumber \\
\frac{V}{1+\hat{m}_V} &=& C_1 \xi_{\perp} (E), \ \ 
\frac{2\hat{m}_V}{X} A_0 = \xi_{\parallel}(E) \Bigl[ C_1
+\Bigl(1-\frac{X}{2} \Bigr) C_2 +C_3 \Bigr], \nonumber \\ 
\frac{1+\hat{m}_V}{X} A_1 &=& C_1 \xi_{\perp} (E), \ \ 
\frac{A_2}{1+\hat{m}_V} = C_1 \xi_{\perp} (E) -\Bigl( C_1
+\frac{X}{2} C_2 +C_3 \Bigr)
\xi_{\parallel} (E).
\end{eqnarray}
These results are the same as those
derived by Bauer et al. \cite{bauer1}, though our basis is different
from theirs. In Ref.~\cite{beneke}, Beneke and Feldmann have
calculated the soft contribution to the form factors using the
large-energy effective theory. As we have stressed, the matching to
the full theory is impossible in this case. However, they judiciously
absorbed the infrared divergences into the nonperturbative parameters
such as $\xi_P$, $\xi_{\perp}$ or $\xi_{\parallel}$ by observing the
Dirac structure of the matrix elements. In the process, the
nonperturbative parameters are defined at each order in
$\alpha_s$. Since we can match the collinear effective theory to the
full theory, we can check their calculations. We find that their
perturbative corrections in Eqs.~(30), (32) and (33) in
Ref.~\cite{beneke} are correct when we compare them with the exact
results in the collinear effective theory. Now we include the
nonperturbative corrections at order $\lambda$, along with the
perturbative correction.  

At order $\lambda$, the matrix elements of $O_i^{\mu}$ for
pseudoscalar mesons are given as
\begin{eqnarray}
\langle P |\overline{\xi}_n \frac{\FMslash{\overline{n}}}{2}
\FMslash{p}_{\perp} \gamma^{\mu} h_v |\overline{B} \rangle &=& a_P (2
v^{\mu} -n^{\mu}), \ \langle P |\overline{\xi}_n
\frac{\FMslash{\overline{n}}}{2} \FMslash{p}_{\perp} \gamma^{\mu}
\gamma_5 h_v |\overline{B} \rangle =0, \nonumber \\ 
\langle P |\overline{\xi}_n \frac{\FMslash{\overline{n}}}{2}
\FMslash{p}_{\perp} v^{\mu} h_v |\overline{B} \rangle &=& a_P v^{\mu},
\ \langle P |\overline{\xi}_n \frac{\FMslash{\overline{n}}}{2}
\FMslash{p}_{\perp} v^{\mu} \gamma_5 h_v |\overline{B} \rangle =0,
\nonumber \\ 
\langle P |\overline{\xi}_n \frac{\FMslash{\overline{n}}}{2}
\FMslash{p}_{\perp} n^{\mu} h_v |\overline{B} \rangle &=& a_P n^{\mu},
\  \langle P |\overline{\xi}_n \frac{\FMslash{\overline{n}}}{2}
\FMslash{p}_{\perp} n^{\mu} \gamma_5 h_v |\overline{B} \rangle =0,
\nonumber \\ 
\langle P |\overline{\xi}_n p_{\perp}^{\mu} h_v |\overline{B} \rangle
&=&  \langle P |\overline{\xi}_n p_{\perp}^{\mu} \gamma_5 h_v
|\overline{B} \rangle =0,
\end{eqnarray}
and for vector mesons, we have
\begin{eqnarray}
\langle V |\overline{\xi}_n \frac{\FMslash{\overline{n}}}{2}
\FMslash{p}_{\perp} \gamma^{\mu} \gamma_5 h_v |\overline{B} \rangle
&=&a_{V2} \epsilon^* \cdot v (2v^{\mu} -n^{\mu}),
\ \  \langle V |\overline{\xi}_n \frac{\FMslash{\overline{n}}}{2}
\FMslash{p}_{\perp} \gamma^{\mu} h_v |\overline{B} \rangle = 0, \ \ 
\nonumber \\ 
\langle V |\overline{\xi}_n \frac{\FMslash{\overline{n}}}{2} 
\FMslash{p}_{\perp} v^{\mu} \gamma_5 h_v |\overline{B} \rangle &=&
-a_{V2} \epsilon^* \cdot v v^{\mu}, \ \ 
\langle V |\overline{\xi}_n \frac{\FMslash{\overline{n}}}{2}
\FMslash{p}_{\perp} v^{\mu} h_v |\overline{B} \rangle = 0, \ \ 
\nonumber \\
\langle V |\overline{\xi}_n \frac{\FMslash{\overline{n}}}{2}
\FMslash{p}_{\perp} n^{\mu} \gamma_5 h_v |\overline{B} \rangle &=&
-a_{V2} \epsilon^* \cdot v n^{\mu}, \ \ 
\langle V |\overline{\xi}_n \frac{\FMslash{\overline{n}}}{2}
\FMslash{p}_{\perp} n^{\mu} h_v |\overline{B} \rangle = 0, \ \ 
\nonumber \\
\langle V |\overline{\xi}_n p_{\perp}^{\mu} \gamma_5 h_v |\overline{B}
\rangle &=& -a_{V1} \Bigl( \epsilon^{*\mu} -(\epsilon^* \cdot v)
n^{\mu} \Bigr), \ \
\langle V |\overline{\xi}_n p_{\perp}^{\mu} h_v |\overline{B} \rangle
=a_{V1} i \epsilon^{\mu \nu \alpha \beta} \epsilon^*_{\nu} n_{\alpha}
v_{\beta}.   
\end{eqnarray}

Finally, for the time-ordered products, we have
\begin{eqnarray}
\langle P | i\int d^4 y T\Bigl\{ \overline{\xi}_n \gamma^{\mu} h_v 
(0)\ \mathcal{L}_1 (y) \Bigr\} |\overline{B}\rangle &=& b_P n^{\mu},
\nonumber \\    
\langle P | i\int d^4 y T\Bigl\{ \overline{\xi}_n v^{\mu} h_v
(0)\ \mathcal{L}_1 (y) \Bigr\} |\overline{B}\rangle &=& b_P v^{\mu},
\nonumber \\ 
\langle P | i\int d^4 y T\Bigl\{ \overline{\xi}_n n^{\mu} h_v
(0)\ \mathcal{L}_1 (y) \Bigr\} |\overline{B}\rangle &=& b_P n^{\mu},
\end{eqnarray}
and the time-ordered products involving the heavy-light currents with
$\gamma_5$ vanish. For the matrix elements of the time-ordered
products for vector mesons, we find
\begin{eqnarray}
\langle V | i\int d^4 y T\Bigl\{ \overline{\xi}_n \gamma^{\mu} h_v 
(0)\ \mathcal{L}_1 (y) \Bigr\} |\overline{B}\rangle &=& b_{V1}
i\epsilon^{\mu \nu \alpha \beta} \epsilon^*_{\nu} n_{\alpha}
v_{\beta},\nonumber \\ 
\langle V | i\int d^4 y T\Bigl\{ \overline{\xi}_n \gamma^{\mu}
\gamma_5 h_v
(0)\ \mathcal{L}_1 (y) \Bigr\} |\overline{B}\rangle &=&b_{V1}
\Bigl(\epsilon^{*\mu} - (\epsilon^* \cdot v) n^{\mu} \Bigr) + b_{V2}
\epsilon^* \cdot v n^{\mu},  \nonumber \\
\langle V | i\int d^4 y T\Bigl\{ \overline{\xi}_n v^{\mu} h_v
(0)\ \mathcal{L}_1 (y) \Bigr\} |\overline{B}\rangle &=& 0 ,\ \ \langle
V | i\int d^4 y T\Bigl\{ \overline{\xi}_n n^{\mu} h_v 
(0)\ \mathcal{L}_1 (y) \Bigr\} |\overline{B}\rangle = 0, \nonumber \\ 
\langle V | i\int d^4 y T\Bigl\{ \overline{\xi}_n v^{\mu} \gamma_5 h_v 
(0)\ \mathcal{L}_1 (y) \Bigr\} |\overline{B}\rangle &=& -b_{V2}
\epsilon^* \cdot v v^{\mu},  \nonumber \\
\langle V | i\int d^4 y T\Bigl\{ \overline{\xi}_n n^{\mu} \gamma_5h_v
(0)\ \mathcal{L}_1 (y) \Bigr\} |\overline{B}\rangle &=& -b_{V2}
\epsilon^* \cdot v n^{\mu}. 
\end{eqnarray}

Combining all these form factors, we obtain in the collinear effective
theory 
\begin{eqnarray}
\langle P | V^{\mu} |\overline{B}\rangle &=& 2E n^{\mu} \Bigl[ (C_1
+C_3) \xi_P +\frac{1}{2E} \Bigl( a_P (-B_1 +B_3) + b_P (A_1+A_3)
\Bigr) \Bigr] 
\nonumber \\
&&+2E v^{\mu} \Bigl[ C_2 \xi_P +\frac{1}{2E} \Bigl( a_P (2B_1 +B_2) +
b_P A_2 \Bigr) \Bigr], \nonumber \\
\langle V|V^{\mu} |\overline{B}\rangle &=& 2E i \epsilon^{\mu \nu
\alpha \beta} \epsilon^*_{\nu} n_{\alpha} v_{\beta} \Bigl[ C_1
\xi_{\perp} +\frac{1}{2E} \Bigl( B_4 a_{V1}+A_1 b_{V1} \Bigr) \Bigr],
\nonumber \\
\langle V|A^{\mu} |\overline{B}\rangle &=& 2E \epsilon^{*\mu} \Bigl[
C_1 \xi_{\perp} + \frac{1}{2E} \Bigl( B_4 a_{V1}+ A_1 b_{V1}\Bigr)
\Bigr]  \nonumber \\ 
&-& 2E(\epsilon^* \cdot v) n^{\mu} \Bigl[C_1 \xi_{\perp} -
(C_1 +C_3) \xi_{\parallel} \nonumber \\
&&+\frac{1}{2E} \Bigl( (B_1 -B_3) a_{V2} +B_4
a_{V1} +A_1 b_{V1} -(A_1+A_3) b_{V2}  \Bigr) \Bigr] \nonumber \\
&+&2E (\epsilon^* \cdot v) v^{\mu} \Bigl[ C_2 \xi_{\parallel}
+\frac{1}{2E} \Bigr( (2B_1+B_2) a_{V2}+ A_2 b_{V2}\Bigr)
\Bigr].  
\end{eqnarray}
From these relations, we can obtain the form factors to order
$\lambda$ and to leading-logarithmic order in $\alpha_s$ as
\begin{eqnarray}
f_+ &=& \Bigl[C_1 +\frac{X}{2} C_2 +C_3 \Bigr] \Bigl[ \xi_P
+\frac{1}{2E} (a_P + b_P) \Bigr]-(2-X) C_1 \frac{a_P}{2E}, \nonumber 
\\   
\frac{f_0}{X} &=& \Bigl[ C_1 + \Bigl(1-\frac{X}{2} \Bigr)
C_2 + C_3 \Bigr] \Bigl[ \xi_P +\frac{1}{2E} (a_P + b_P)\Bigr]
-XC_1 \frac{a_P}{2E},\nonumber \\
\frac{2\hat{m}_V}{X} A_0 &=& \Bigl[ C_1 +\Bigl(1-
\frac{X}{2} \Bigr) C_2 +C_3 \Bigr] \Bigl[ \xi_{\parallel}
+\frac{1}{2E} (a_{V2} + b_{V2})\Bigr] -XC_1 \frac{a_{V2}}{2E},
\nonumber \\
\frac{1+\hat{m}_V}{X} A_1 &=&\frac{V}{1+\hat{m}_V}=
C_1 \Bigl(\xi_{\perp}
+\frac{b_{V1}}{2E} \Bigr) +C_3 \frac{a_{V1}}{E},  \nonumber \\ 
\frac{A_2}{1+\hat{m}_V} &=& C_1 \Bigl( \xi_{\perp} +\frac{b_{V1}}{2E}
\Bigr) + C_3 \frac{a_{V1}}{E} - \Bigl[ C_1 +\frac{X}{2} C_2 +C_3
\Bigr] \Bigl[ \xi_{\parallel} 
+\frac{1}{2E} (a_{V2} + b_{V2})\Bigr] \nonumber \\
&&+(2-X) C_1 \frac{a_{V2}}{2E}. 
\label{formlam} 
\end{eqnarray}
Here we keep $\hat{m}_V$ explicitly even though $\hat{m}_V \sim
\Lambda_{\mathrm{QCD}}/m_b \sim \lambda^2$ in our power counting. It 
is because meson masses are inserted in the definition of form factors
in Eq.~(\ref{parform}) without regard to the power counting in the
collinear effective theory. However, we neglect the terms proportional
to the mass squared of the light meson compared to $M^2$. And
we use the relations among the Wilson coefficients to express the
result in terms of $C_i$ only.

At leading order in $\lambda$, there are three 
unknown nonperturbative parameters $\xi_P(E)$, $\xi_{\perp}(E)$, and 
$\xi_{\parallel}(E)$. These are dimensionless functions. While the
Isgur-Wise function in HQET is normalized to one at maximal momentum
transfer due to the heavy quark symmetry, there is no constraint in
the normalization of these unknown parameters \cite{beneke}. 
At order $\lambda$, there are six additional 
nonperturbative parameters: $a_P(E)$, $a_{V1}(E)$, $a_{V2} (E)$, $b_P
(E)$, $b_{V1}(E)$, and $b_{V2} (E)$. In our convention, all these
parameters have mass dimension, for example,
\begin{equation}
\frac{a_P}{E}\sim \lambda \sim
\sqrt{\frac{\Lambda_{\mathrm{QCD}}}{m_b}},
\end{equation}
where the last relation comes from the kinematics. The remaining five
unknown parameters are of the same order in $\lambda$. Therefore
Eq.~(\ref{formlam}) is our result for the form factors to order
$\sqrt{\Lambda_{\mathrm{QCD}}/m_b}$. 

There are interesting relations among the form factors in the
effective theory. At zeroth order in $\lambda$ and $\alpha_s$, those
relations are given by 
\begin{eqnarray}
f_+ &=& \frac{f_0}{X}(=\xi_P), \ \ \frac{V}{1+\hat{m}_V} =
\frac{1+\hat{m}_V}{X} A_1(=\xi_{\perp}),\nonumber \\ 
\frac{2\hat{m}_V}{X} A_0 &=& \frac{1+\hat{m}_V}{X} A_1 -(1-\hat{m}_V)
A_2(=\xi_{\parallel}).
\label{formlead}
\end{eqnarray}
These relations are modified at order $\lambda$ and at
leading-logarithmic order in $\alpha_s$ as
\begin{eqnarray}
f_+ -\frac{f_0}{X} &=& -(1-X) \Bigl[ C_2 \Bigl( \xi_P +\frac{1}{2E}
(a_P + b_P) \Bigr) +\frac{1}{E} C_1 a_P\Bigr], \nonumber \\
\frac{V}{1+\hat{m}_V} &=&
\frac{1+\hat{m}_V}{X} A_1,\nonumber \\ 
\frac{2\hat{m}_V}{X} A_0&=&\frac{1+\hat{m}_V}{X} A_1 -(1-\hat{m}_V)
A_2 \nonumber \\
&&+ (1-X) C_2 \Bigl[ \xi_{\parallel} +\frac{1}{2E}(a_{V2} + b_{V2})
\Bigr] +(1-X)C_1 \frac{a_{V2}}{E}.
\label{formdiff}
\end{eqnarray}
Note that the second relation in Eq.~(\ref{formlead}) still holds to
order $\lambda$ and at leading-logarithmic order in $\alpha_s$. And
the tree-level results hold only in the limit $X\rightarrow 1$.

\section{\label{conc}Conclusion}
We have shown that heavy meson decays in which light mesons are
emitted with large energy can be consistently described by the
collinear effective theory combined with the HQET. And we can obtain a 
systematic expansion of the effective Lagrangian in powers of
$\lambda$. Heavy-light currents can also be
expanded consistently in powers of $\lambda$, and the Wilson
coefficients of various operators in the effective theory can be
computed by matching the effective theory to the full theory. It is
crucial to note that the collinear effective theory reproduces the
infrared behavior of the full theory by including the effects of
collinear gluons. 

There is a reparameterization invariance in the collinear effective
theory, in which a slight change of the light-cone direction $n^{\mu}$
can be compensated by a change of $p_{\perp}$ to make the physics
invariant under this transformation. If we also require that the
theory be invariant under collinear gauge transformations, we can
prove that the effective Lagrangian $\mathcal{L}_1$ at order $\lambda$
is not renormalized. This reparameterization invariance is also useful
in deriving the operators of order $\lambda$ from the operators of
order $\lambda^0$. The Wilson coefficients and the anomalous
dimensions can be obtained from the operators which are related by the 
reparameterization invariance. The reparameterization invariance and
the collinear gauge invariance put a serious
constraint in the structure of heavy-light currents in the collinear
effective theory.

The development of the collinear effective theory casts a renewed
view on heavy quark decays in which light quarks are emitted with
large energy. Bauer et al. \cite{bauer4} have considered nonleptonic
decays using the collinear effective theory, and found that the decay 
$B\rightarrow D\pi$ is factorized in the heavy quark limit to all
orders in $\alpha_s$. It will be interesting to look into nonleptonic
decays of $B$ mesons in the context of the collinear effective
theory including higher-order corrections in $\lambda$. 

What we have not considered here is hard spectator effects, in which
spectator quarks interact with the energetic quark through hard
gluons. As Beneke et al. \cite{beneke} pointed out, this contribution
can be as important as the soft contribution to the form factors. If
we can analyze the hard spectator contribution also in the scheme of
the collinear effective theory, we will have a better understanding of
form factors in this kinematic region. This is the next subject to be
developed. 

\begin{acknowledgments}
The authors are supported by the
Ministry of Education Grants KRF-99-042-D00034, KRF-2000-015-DP0067,
and Hacksim BK21 Project.
\end{acknowledgments}

\appendix*

\section{\label{appen}Renormalization of $\mathcal{L}_1$ at order
$\alpha_s$} 
\begin{figure}[t]
\hspace{5.0cm}
\begin{center}
\epsfig{file=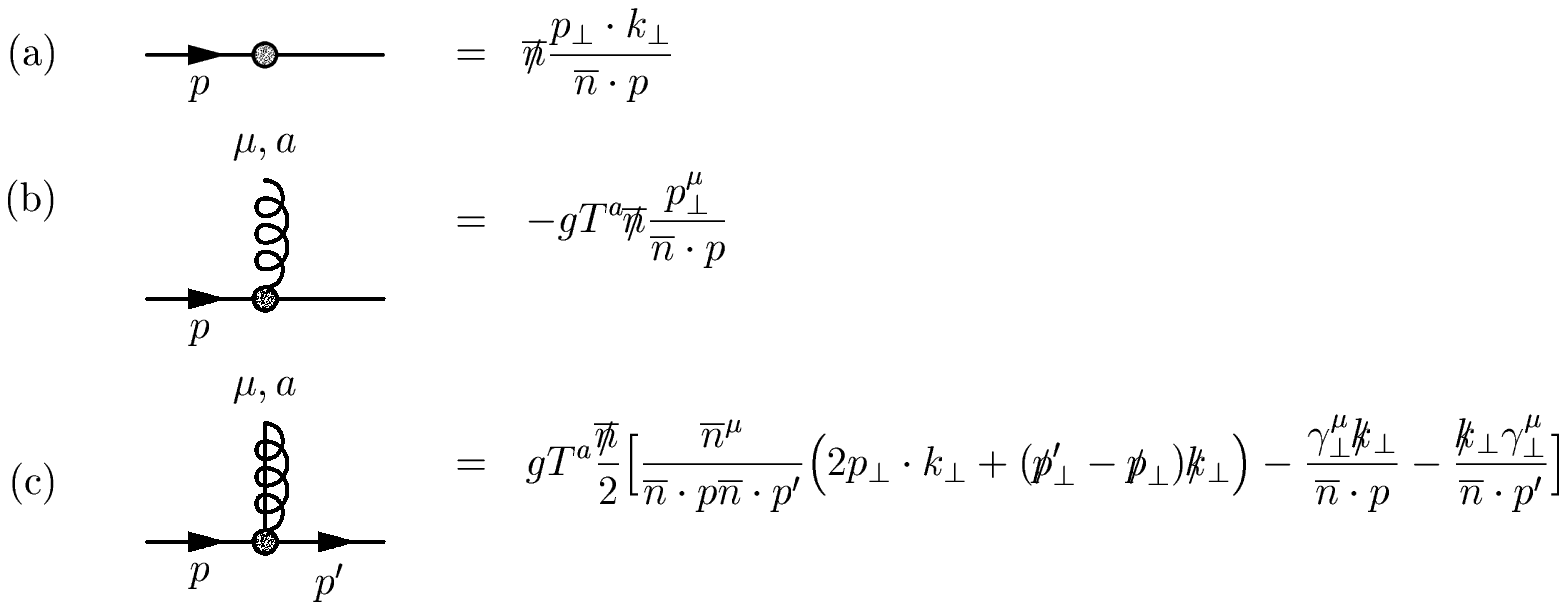}
\end{center}
\caption{Feynman rules for the effective Lagrangian $\mathcal{L}_1$ to
order $g$: (a) collinear quark without an external gluon, (b)
collinear quark interaction with a soft gluon, and (c) collinear quark
interaction with a collinear gluon, and $k^{\mu}$ denotes residual
momentum of order $\lambda^2$.}
\label{lone}
\end{figure}

In this Appendix, we show explicitly that the effective
Lagrangian $\mathcal{L}_1$ at order $\lambda$ in Eq.~(\ref{eftlag}) is
not renormalized at one loop. 
The Feynman rules for the Lagrangian $\mathcal{L}_1$ to order $g$ is
shown in Fig.~\ref{lone}. The derivative is of
order $\lambda^2$, and it is replaced by the residual momentum $k$ in
momentum space.
We will concentrate on the first term in $\mathcal{L}_1$, which is of
the form 
\begin{equation}
O_1 = \overline{\xi}_n 
\frac{\FMslash{p}_{\perp} i\FMslash{\partial}_{\perp}
+i\FMslash{\partial}_{\perp}\FMslash{p}_{\perp} }{\overline{n} \cdot
p} \frac{\FMslash{\overline{n}}}{2} \xi_n.
\end{equation}
which is shown in Fig.~\ref{lone} (a). Other terms in
$\mathcal{L}_1$  contribute 
to the renormalization of $O_1$ at order $\lambda$ along with the
radiative corrections of $O_1$. In order to show that $\mathcal{L}_1$
is not renormalized, we have to consider all the radiative corrections
for the operators shown in Fig.~\ref{lone}. However, we will
concentrate on the renormalization of $O_1$, since other terms have
the same renormalization behavior as $O_1$ at leading logarithmic
order.

\begin{figure}[b]
%\hspace{-5.5cm}
\epsfig{file=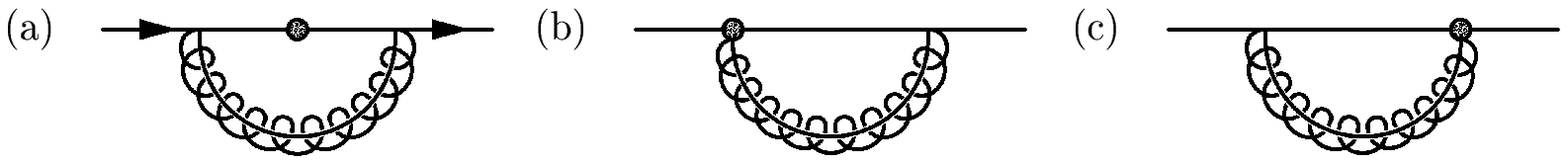}
\caption{Feynman diagrams for the renormalization of $\mathcal{L}_1$
at one loop.}
\label{renl1}
\end{figure}

The Feynman diagrams to renormalize $O_1$ are shown in
Fig.~\ref{renl1}. And the corresponding diagrams with a soft gluon
exchange vanish due to the vertex structure. All the 
diagrams in Fig.~\ref{renl1} are zero using dimensional regularization
for on-shell 
external states, and the coefficient of $O_1$ is given by the
tree-level value. In order to see the renormalization group behavior, 
we have to extract the ultraviolet divergent part by 
putting the external quark off the mass shell by $p^2 =
p_{\perp}^2$. We will show only the ultraviolet divergent
parts here. Calculating the Feynman diagram in Fig.~\ref{renl1} (a),
(b) and (c), we obtain  
\begin{equation}
M_a = \frac{\alpha_s C_F}{4\pi} \frac{1}{\epsilon} O_1, \ \ 
M_b = -\frac{\alpha_s C_F}{4\pi} \frac{3}{\epsilon} O_1,\ \ 
M_c = \frac{\alpha_s C_F}{4\pi} \frac{3}{\epsilon} O_1,
\end{equation}
respectively. Therefore the sum of all the diagrams is given by
\begin{equation}
M=M_a + M_b + M_c = \frac{\alpha_s C_F}{4\pi}
\frac{1}{\epsilon} O_1. 
\end{equation}
When we add the wave function renormalization to this amplitude, 
the ultraviolet divergences cancels, and the anomalous dimension of
$O_1$ is zero. Therefore we have shown that the operator $O_1$ is not
renormalized at order $\alpha_s$ explicitly. In fact, we have to
consider one-loop corrections to the remaining operators in
$\mathcal{L}_1$. But no other operators are renormalized though we do
not show them here. As a result, the Wilson coefficients $A_k$
of the time-ordered products in Eq.~(\ref{ai}) come from the Wilson
coefficients of the operators $J^{\mu}_i$ alone and not from
$\mathcal{L}_1$.

\end{document}